\documentclass[useAMS,usenatbib]{mn2e}
\usepackage{graphicx, times, ulem}
\usepackage{graphicx}
\usepackage{subfigure}
\usepackage{natbib}
\usepackage{color,hyperref,topcapt}
\usepackage{amssymb,amsmath}
\usepackage{bm}
\usepackage{gensymb}

\title{Dancing with the Stars: Formation of the Fomalhaut triple system and its effect on the debris disks}
\author[Shannon et. al.]{Andrew Shannon, Cathie Clarke, \& Mark Wyatt \\
Institute of Astronomy, University of Cambridge, Madingley Road, Cambridge, United Kingdom of Great Britain and Northern Ireland, CB3 0HA}

\begin{document}

\pagerange{\pageref{firstpage}--\pageref{lastpage}} \pubyear{2014}

\maketitle

\begin{abstract}
Fomalhaut is a triple system, with all components widely separated ($\sim 10^5$~au).  Such widely separated binaries are thought to form during cluster dissolution, but that process is unlikely to form such a triple system.  We explore an alternative scenario, where A and C form as a tighter binary from a single molecular cloud core (with semimajor axis $\sim 10^4$~au), and B is captured during cluster dispersal.  We use N-body simulations augmented with the Galactic tidal forces to show that such a system naturally evolves into a Fomalhaut-like system in about half of cases, on a timescale compatible with the age of Fomalhaut.  From initial non-interacting orbits, Galactic tides drive cycles in B's eccentricity that lead to a close encounter with C.  After several close encounters, typically lasting tens of millions of years, one of the stars is ejected.  The Fomalhaut-like case with both components at large separations is almost invariably a precursor to the ejection of one component, most commonly Fomalhaut C.  By including circumstellar debris in a subset of the simulations, we also show that such an evolution usually does not disrupt the coherently eccentric debris disk around Fomalhaut A, and in some cases can even produce such a disk.  We also find that the final eccentricity of the disk around A and the disk around C are correlated, which may indicate that the dynamics of the three stars stirred C's disk, explaining its unusual brightness.
\end{abstract}


\section{Introduction}

\label{section:intro}

Fomalhaut has been suspected to be part of a widely separated binary star system for some time now \citep{1938AJ.....47..115L}.  Recent analysis has confirmed that the K4V star TW PsA has both a similar proper motion and radial velocity to Fomalhaut, such that it is not an interloping field star, but forms a binary star system with Fomalhaut.  The pair have a three-dimensional separation of $5.74^{+0.04}_{-0.03}\times 10^4 \rm{au}$.   Combining isochronal, rotational, X-ray, and lithium ages for the pair, the system is constrained to have an age of $440 \pm 40~\rm{Myrs}$, the mass of Fomalhaut A to be $1.92 \pm 0.02 M_{\oplus}$, and the mass of Fomalhaut B to be $0.73^{+0.02}_{-0.01}M_{\oplus}$ \citep{2012ApJ...754L..20M}.  More recently, a third member of the Fomalhaut system has been recognised.  The M4V star LP 876-10, or Fomalhaut C, is a $0.18 \pm 0.02 M_{\odot}$~star with a three dimensional separation from Fomalhaut A of $1.58^{+0.02}_{-0.01}\times 10^5 \rm{au}$~\citep{2013AJ....146..154M}, which also has a common proper motion with Fomalhaut A and B.  The uncertainty in the measured velocities of the three stars is roughly half a kilometre per second, much smaller than the relative velocities expected between field stars, allowing the inference that the system is bound.  However, the escape speed for B is about a quarter kilometre per second, and C's smaller still, so little can be said about the orbital configuration.

The dynamics of the Fomalhaut system are particularly interesting as Fomalhaut A is known to harbour a debris disk \citep{1986ASSL..124...61G}, which has a coherent eccentricity of $0.11 \pm 0.01$ \citep{2005Natur.435.1067K}.  The origin of the debris disk's eccentricity is not known.   It has been suggested it may result from the action of one or more shepherding planet(s) \citep{2006MNRAS.372L..14Q,2009ApJ...693..734C,2012ApJ...750L..21B}, although other possible origins have been advanced \citep[e.g.,][]{2013Natur.499..184L}.  Fomalhaut A is also accompanied by a point-like object dubbed 'Fomalhaut b' \citep{2008Sci...322.1345K}.  Non-detection of Fomalhaut b at thermal wavelengths led \citet{2008Sci...322.1345K} to the suggestion that it may be the dust cloud ejected in a collision between planetesimals, or a circumplanetary ring, and \citet{2011MNRAS.412.2137K} modelled it as a circumplanetary dust cloud created by the collisions of irregular satellites.  Continuing observation has allowed for improved fits to the orbit of Fomalhaut b, which have revealed that it is not a shepherding planet, but has an eccentricity of $\sim 0.8$, and a semimajor axis similar to the dust belt bodies \citep{2013ApJ...775...56K,2014A&A...561A..43B}.  This eccentric orbit precludes Fomalhaut b having a mass significantly above $10 M_{\oplus}$, unless it was recently scattered to its highly eccentric orbit, as a more massive planet would perturb the disk to disruption \citep{2014A&A...561A..43B,2014MNRAS.tmp..149T}.  The origin of the eccentricities of the dust belt's eccentricity, Fomalhaut b's eccentricity, and the nature of Fomalhaut b all remain open questions.

Moreover, Fomalhaut C has also been discovered to have a dust disk \citep{2014MNRAS.438L..96K}.  Debris disks are rare around M stars \citep{2007ApJ...667..527G}, and thus in the Fomalhaut system, which offers the possibility of additional dynamical constraints, the existence of a debris disk about an M star is of particular interest.

The formation of such a wide triple system presents some problems. The separations of both components are larger than the typical sizes of star forming cores and thus, given that such cores rotate at far less than break-up velocity \citep{1993ApJ...406..528G}, the system contains far too much angular momentum to have been created by conventional core fragmentation.  The unfolding of triple stellar systems suggested by \citep{2012Natur.492..221R} appears to be inapplicable to Fomalhaut, which lacks an inner binary.  Capture during cluster dispersal has also been suggested as a mechanism for wide binary formation \citep{2010MNRAS.404.1835K,2011MNRAS.415.1179M}, but the low probability of capture makes the formation of a triple system by this method highly unlikely.  The fraction of field stars with wide binaries is $\sim 10^{-2}$, \citep{2010AJ....139.2566D}, suggesting that the chance of forming a cluster capture binary $\sim 10^{-2}$, and thus forming a triple this way should have odds $\sim 10^{-4}$.  In comparison, the chance of forming a $10^4$~au binary is $\sim 10^{-1}$, \citep{2014MNRAS.437.1216D}.  Thus we are motivated to consider an alternative scenario, where AC form as a tighter binary from a single core, and C is moved outwards by interactions with B, which is captured as a wide binary.  We describe our numerical method in \textsection \ref{section:method}, and present the results of our simulations in \textsection \ref{section:results}.  We perform a subset of simulations with debris disks encircling Fomalhaut A and C in \textsection \ref{section:disk} to assess the compatibility of this scenario with the observed disks.

\section{Numerical Method}

\label{section:method}

We perform simulations with the radau integrator in the MERCURY suite of N-body integrators \citep{1999MNRAS.304..793C}.  We augment this with the Galactic tidal prescription from \citet{2013MNRAS.430..403V}, with parameters appropriate for the solar neighbourhood.  As described in \textsection \ref{section:intro} we consider the scenario where A and C formed as a binary within a single core of the molecular cloud, and Fomalhaut B was bound to the AC pair during cluster dispersal.  Today, the total energy of the system is 
\begin{equation}
 E_{total} \gtrsim -\frac{GM_{A}M_{B}}{R_{AB}} - \frac{GM_{A}M_C}{R_{AC}}
\end{equation}
If the system formed with $B$~at large separations with a low velocity, such that its contribution to the total energy can be neglected, the initial energy was approximately 
\begin{equation}
 E_0 \approx -\frac{GM_AM_C}{2a_{AC,0}}
\end{equation}
Thus, equating the two energies, the primordial AC binary must have had a semimajor axis of at least
\begin{equation}
 a_{AC,0} \gtrsim \frac{M_CR_{AC}R_{AB}}{2\left(M_BR_{AC}+M_CR_{AB}\right)} \approx 6400 \rm{au}
 \label{eq:ainit}
\end{equation}
which is compatible with the $\sim 10^4~\rm{au}$~size of molecular cores \citep{1983ApJ...266..309M,2011psf..book.....B,2013MNRAS.432.3288S}.  Considering the corresponding expression for the case of a primordial AB binary from a single core with C loosely bound during cluster dispersal shows that the AB semimajor axis could not have been substantially increased from its primordial value, precluding such a scenario. 

For the initial AC binary, we choose eccentricities randomly from a flat distribution from $0 \rightarrow 1$, similar to what is observed for solar-type binaries with measured orbits greater than 12 days \citep{2010ApJS..190....1R}.  Given the minimum set by equation \eqref{eq:ainit}, and that the largest molecular cores are slightly larger than $10^4 \rm{au}$, we choose the initial semimajor axis from a flat distribution in $log{\left(a\right)}$~extending over a factor $e$~from 5000 to 13600 au.  The exact limits are slightly arbitrary, but cover the range of interest without considering unphysical values.  The orbit of the AC binary is oriented randomly with respect to the Galactic plane.  We place B at a randomly chosen location within the Hill sphere of the ABC system around the galaxy, which in the Galactic tidal potential is an ellipsoid with a maximum radius of $4.09 \times 10^5 \rm{au}$, or 1.99 pc \citep{2013MNRAS.430..403V}.   We assign B a random velocity given by 
\begin{equation}
 f_{V}\left(V\right)dV = \frac{1}{2\sqrt{\pi}\sigma^3}e^{-\frac{V^2}{4\sigma^2}}V^2dV
\end{equation}
with $\sigma = 1.9 \rm{km}\rm{s}^{-1}$, a plausible value for a dissolving cluster \citep{2010MNRAS.404.1835K}.  If the chosen $V$~has B energetically unbound, we regenerate the system\footnote{In practice, most generations give an unbound B, so the precise choice of $\sigma$ is not very significant, as it is the requirement the system be energetically bound which determines the initial orbit of B.}.  This yields an approximately thermal distribution of eccentricities for B's orbit - $f\left(e\right) \sim e$.  As the age of Fomalhaut is measured to be $440 \pm 40$ Myrs, simulations are run for 500 Myrs.  Stars are removed from the simulation if they are more than 410 000 au (1.99 pc) from Fomalhaut A, appropriate for our Galactic tidal model \citep{2013MNRAS.430..403V}.

\section{Results}

\label{section:results}

\begin{figure}
  \centering
  {\includegraphics[width=0.50\textwidth,trim = 120 50 100 50, clip]{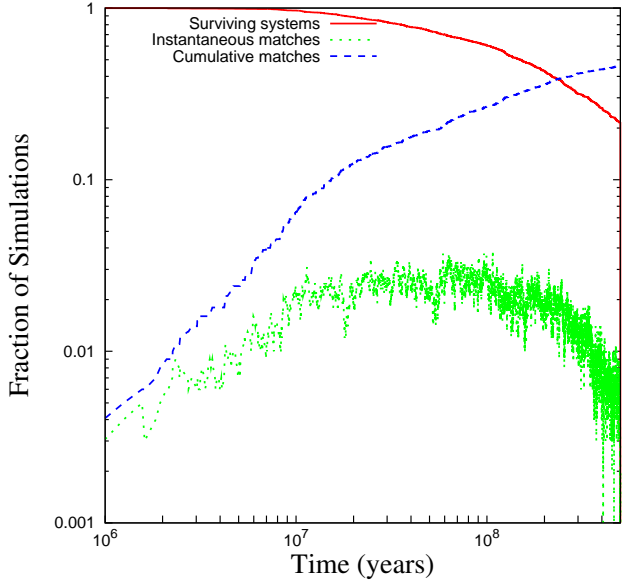}}
  \caption{Evolution of the population, with the total number of systems still having all three stars (red solid line), the instantaneously matching systems (green dotted line), and the total number of systems that have ever matched (blue dashed line).  At 500 Myrs, 459 systems have ever been a match for the Fomalhaut system.  In addition, 214 systems are still active, 191 of which have never been a match; these reside in roughly their original, unperturbed state. }
\label{fig:active}
\end{figure}

\begin{figure}
  \centering
  {\includegraphics[width=0.50\textwidth,trim = 120 50 100 50, clip]{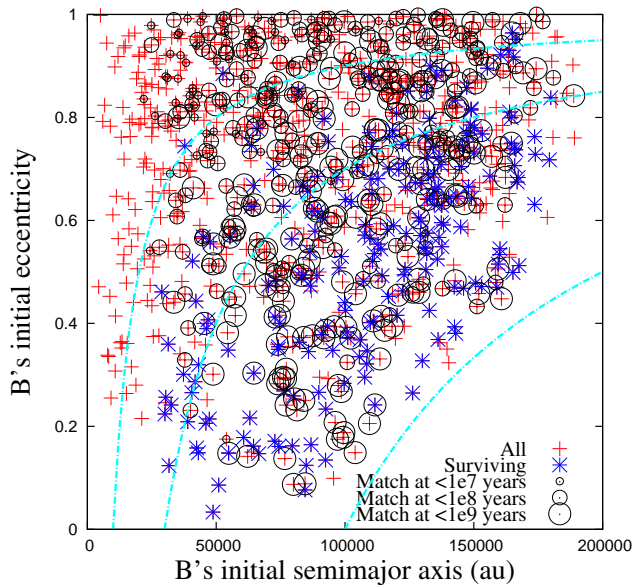}}
  \caption{Initial semimajor axis and eccentricity of B's orbit about AC, showing if and when the system became a match, as well as those systems still active at 500 Myrs.  Systems with higher initial eccentricity, or lower initial semimajor axis, tend to become matches earlier.  The surviving systems lie at high semimajor axis and low eccentricity, suggesting some may evolve into Fomalhaut-like systems at later times.  A similar plot of C's initial orbit about A shows no significant trends.  Dash dotted lines are overplotted at initial pericentres of $10^4$, $3\times 10^4$, and $10^5$~au for visualisation.}
\label{fig:binit}
\end{figure}

We perform a total of $1000$ simulations.  We define a simulation to be a match for the Fomalhaut system if the $AC$ separation and $AB$ separation are simultaneously between $0.5$~and $1.5$~times their current values.  Over $500$~Myrs~$459$~of the simulations are ever a match for the Fomalhaut system (figure \ref{fig:active}).  At 500 Myrs, the end of our simulations, 214 systems still retain all three stars\footnote{Most other simulations ejected B or C, although twelve ended in star-star collisions.  We assumed radii of $1.39 R_{\odot}$, $1.01 R_{\odot}$, and $0.63 R_{\odot}$~to calculate collisions for Fomalhaut A, B, and C respectively.}, 191 of which have never been matches.  Figure \ref{fig:binit} shows that the simulations where Fomalhaut B begins with lower semimajor axes and higher eccentricities were the first to go unstable, and the trend is for systems with higher semimajor axes and lower eccentricities to become unstable, and possibly pass through a Fomalhaut-like phase, at later times (figure \ref{fig:binit}).  With a $46\%$~match rate and $19\%$~of systems remaining to become unstable at a later date, we expect $55\% - 60\%$~of systems will eventually pass through a Fomalhaut-like phase.  A similar analysis of the initial orbital parameters of the AC binary reveals no trends; the timing of the instability appears to be set only by the time taken to lower the pericentre of B's orbit enough to allow a close encounter with C.

\begin{figure}
  \centering
  {\includegraphics[width=0.50\textwidth,trim = 120 50 100 50, clip]{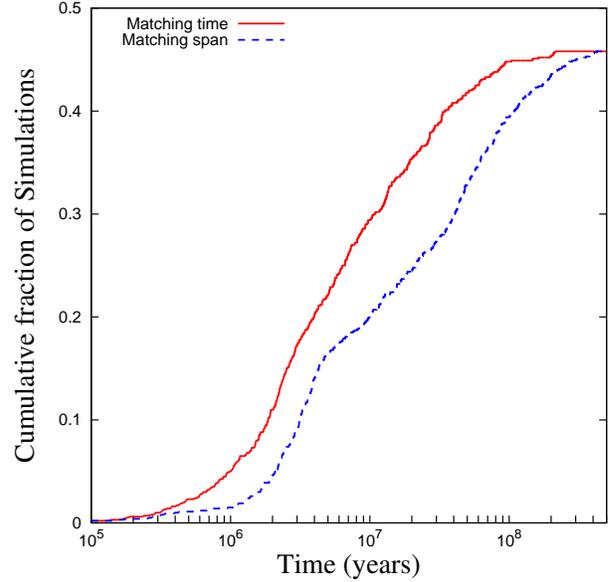}}
  \caption{Cumulative number of simulations by the total amount of time for which they are a match to the Fomalhaut system (solid red line), and by the span of time between when they first become a match, and when they are last are a match (blue dotted line).}
\label{fig:lifetimes}
\end{figure}

The systems that are ever a match spend a mean of $17.2$~Myrs matching, and a median of $5.5$~Myrs matching (see figure \ref{fig:lifetimes}).  We match the simulations to the {\it instantaneous} separations in Fomalhaut (since its orbital trajectories are unknown) and thus simulated systems may move in and out of matching our criterion as the stars progress around their orbits without significant change in their orbital elements.  Thus, our matching criteria may be somewhat too strict.  If we instead measure the time between the first and last occasions the simulation is a match to the Fomalhaut system, the mean matching span is $47.3$~Myrs, and the median is $15.6$~Myrs (see figure \ref{fig:lifetimes}), which corresponds to a few orbits of C around the AB pair. 


As evidenced by the outcome that only 23 of the 459 systems which were matches for Fomalhaut persisted to 500 Myrs, the matching state is a temporary one, which is typically followed by the ejection of one of the stars, most often C\footnote{A total of 617 Cs were lost and 175 Bs, including simulations which were never a match to Fomalhaut.}.  This is perhaps not surprising - after B and C exchange places as the outer and inner binaries respectively, their orbits will remain crossing. 

For B and C to be removed from crossing orbits before one is ejected would require the Galactic tide to raise the pericentre of C by an amount $\Delta q \sim q$.  This takes a time \citep{1987AJ.....94.1330D}
\begin{equation}
 \label{eq:tautides}
 \tau_{tides} = \frac{M_*}{5 \pi^2 \rho_0} a^{-3} \sqrt{\frac{q}{a}} P
\end{equation}
 where $M_* = M_A+M_B+M_C$, the mass of the three stars in this case, $\rho_0$~is the mass density of stars in the local Galactic volume $\sim 2 \times 10^{-17} M_{\odot} \rm{au}^{-3}$~(or $0.2 M_{\odot} \rm{pc}^{-3}$), and $a$, $q$, and $P$~are the semimajor axis, pericentre distance, and period of the outermost star respectively.  The Hill sphere of B about A, is $\sim 0.5$~times the A-B separation, and thus C will be strongly scattered during each pericentre passage.  Substituting the current system configuration into equation \ref{eq:tautides}, assuming B and C have moderate eccentricities, gives a tidal time of order the orbital period of C.  On figure \ref{fig:lifetimes}, we see that systems which become a match for Fomalhaut are typically destroyed after a few orbits of the outer body.  Although the timescale for tides to raise C's pericentre is is comparable to the timescale on which we expect a strong scattering to occur, we find that in only $4-5\%$~of the simulations does the Galactic tide decouple B's and C's orbit before one is ejected.  The chance of observing such a decoupled system is higher than this, as the decoupling of the orbits makes the system longer lived.  These total matching spans of these systems is $\sim 20\%$ of the total matching spans of all systems, and as these simulations are only uncoupled for part of their matching span, we conclude that if the scenario we propose here is correct, we are unlikely ($< 20\%$) to find B and C on widely separated, noninteracting orbits, once their velocities are measured with precision.

\begin{figure}
  \centering
  {\includegraphics[width=0.50\textwidth,trim = 120 50 100 50, clip]{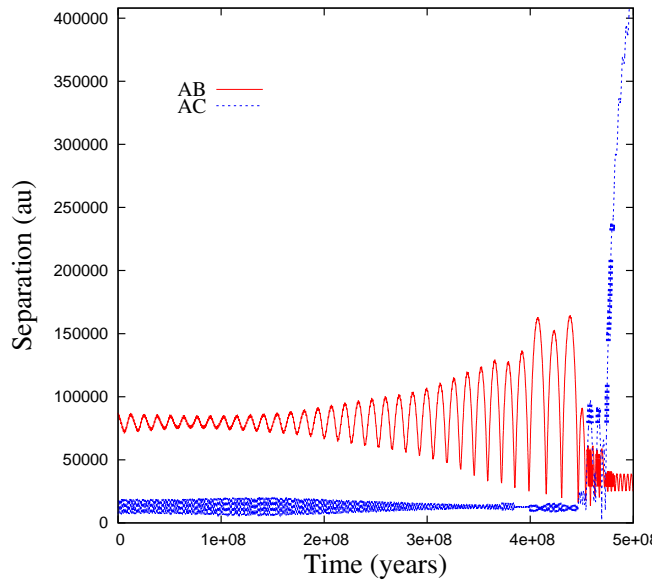}}
  \caption{Evolution of a typical system.  Fomalhauts B and C undergo several scattering encounters, during which the system resembles the actual Fomalhaut system.  The thicker line segments denote where the system is identified as a match to Fomalhaut with our matching criteria.}
\label{fig:example}
\end{figure}

We plot an example of a typical evolution in figure \ref{fig:example}.  This system was selected as an example because  the time it spends as a match to Fomalhaut and the interval been the first and last matches both lie between the median and mean of our distribution and the system is a match at roughly the age of Fomalhaut.
Fomalhaut B begins on a nearly circular orbit, but is driven to higher eccentricity by the action of Galactic tides.  At around 380 Myrs, B and C begin to interact by close encounters, leading to an exchange at 450 Myrs.  After a few close encounters with B, Fomalhaut C is ejected from the system.  The decrease in C's eccentricity between 200 and 350 Myrs is not caused by Galactic tides; rather it is caused by secular interactions between B and C.  As the Galactic tide moves B to higher eccentricity, the secular interaction timescale drops from $\sim 700$~Myrs to $\sim 70$~Myrs \citep[see, e.g.,][]{2000ApJ...535..385F}.  Details such as whether the secular interactions cause an increase or decrease in eccentricity vary from simulation to simulation, but do not impact the overall character of the evolution.

\section{Debris disks}

\label{section:disk}

The presence of a coherently eccentric debris disk around Fomalhaut A provides an additional constraint on the system's history.  A detailed study of how much dispersion is compatible with the observations of Fomalhaut's debris disk has not been performed (and is beyond the scope of this work), but modelling of the inner edge favours an inclination dispersion of $\sim 2{\degree}$, and the observed sharp inner edge of the debris suggests an eccentricity dispersion much less than the ring's value of $0.11\pm 0.01$ \citep{2005Natur.435.1067K,2012ApJ...750L..21B}.  Similarly, the dispersion in the longitude of pericentre must be small for the disk as a whole to be eccentric.  Close encounters between Fomalhaut A and one or both of the other stars can disrupt the disk.  In addition, the stellar dynamics may have implications for the debris disk around Fomalhaut C, as debris disks are generally rare around M stars \citep{2009A&A...506.1455L}. 

To investigate this, we perform a smaller sample of 50 simulations in which we place a circular\footnote{Or rather, the orbits would be circular in the absence of the other two stars; for A's disk, C's gravitational tidal field produces typical initial eccentricities of $10^{-4}$~to $10^{-7}$, while around C, A's tidal field typically causes initial eccentricities of $10^{-3}$~to $10^{-6}$} disk of 100 test particles around Fomalhaut A, with semimajor axis distributed randomly in $a$~from 127 au to 143 au \citep{2012ApJ...750L..21B}.  The initial conditions for the stellar orbits are the same as in section \ref{section:method}.  The disk is oriented randomly with respect to all system components, as well as the Galactic potential.  \citet{2014MNRAS.438L..96K}~found that Fomalhaut C also has a detectable debris disk.  As debris disks are rarely detected around M stars, they speculated that the presence of a debris disk around Fomalhaut A and another around Fomalhaut C may be related.  To evaluate the plausibility of this suggestion, we also included a disk of 100 test particles around Fomalhaut C in these 50 simulations.  Based on constraints from the allowed temperature range of the disk detected around C, as well as size constraints from the disk being unresolved in Herschel images, we spread the disk around C with semimajor axis from 10 au to 40 au, roughly the maximum allowable range.  This disk, too, is oriented randomly with respect to all other system components, including A's disk.  

\begin{figure}
  \centering
  {\includegraphics[width=0.50\textwidth,trim = 120 50 100 50, clip]{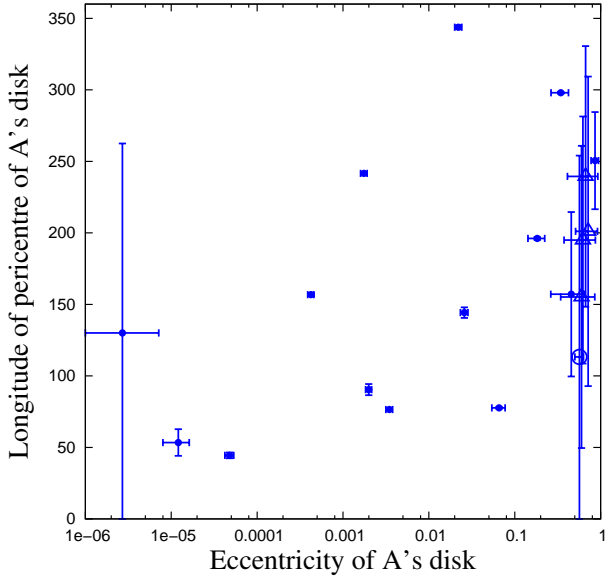}}
  \caption{The mean eccentricty vs the mean longitude of pericentre for all the disk particles immediately after the last occasion the system was identified as a match for the Fomalhaut system.  The error bars show the standard deviation among the disk particles.  Disks begin with a typical eccentricity of $\mathcal{O}\left(10^{-6}\right)$ and average apsidal alignment of $\pm 55 \degree$.  In most cases, the disk eccentricity is raised while the apsides are aligned.  A few disks are significantly disrupted, with high eccentricity $e \gtrsim 0.5$~and no apsidal alignment, in fiv cases including the loss of disk particles (the open symbols).  Thus, in 12 of the 19 cases the disk is preserved in a way compatible with the observed debris disk around Fomalhaut.  Five cases develop a coherent eccentricity with a factor of $5$~of Fomalhaut's $e = 0.11 \pm 0.01$, including two with more eccentric disks. }
\label{fig:disks}
\end{figure}

In this sample, 19 of the systems become a match for the Fomalhaut system over 500 Myrs of evolution.  In figure \ref{fig:disks}, we plot the state of A's disk after the last instance the simulation matches the Fomalhaut system.  Disks around A begin with a typical eccentricity of $\mathcal{O}\left(10^{-6}\right)$ and average apsidal alignment of $\pm 55 \degree$, which we measure using the standard deviation of the longitudes of pericentre of the disk particles.  As the disks rise in eccentricity, they become apsidally aligned, as the eccentricity rise is driven by secular interactions with C, and the disk particles span a small range of initial semimajor axes.  Close encounters can destroy the alignment, and raise eccentricities to high values, which occurs in 7 of the 19 cases (all with $e \gtrsim 0.5$).  Five of the seven disrupted disks also lost disk particles. 


Of particular note is that almost all of the undisrupted disks generate a coherent eccentricity.  Although those cases with eccentricities of $\lesssim 10^{-2}$ are perhaps not evocative of Fomalhaut, they still represent eccentricity increases of 1-3 orders of magnitude from the initial values; such systems would require an internal mechanism to drive the eccentricity of A's disk to the high value seen today. Five disks develop coherent eccentricities between $0.02$~and $0.5$, evocative of Fomalhaut's disk, including two with higher eccentricies of $0.34 \pm 0.08$~and $0.18 \pm 0.04$, showing that the disk eccentricity can coherently rise as high as the value seen in Fomalhaut's disk today.  The coherent eccentricity rising could similarly have produced aligned shepherding planets, as were suggested by \citet{2012ApJ...750L..21B} to explain the ring's morphology.  Previous simulations of stellar fly-bys of debris disks have shown them capable of producing eccentric rings \citep{2001MNRAS.323..402L}, but this had been rejected as an explanation for the eccentricity of Fomalhaut's ring as it was thought that repeated pericentre passages would destroy the coherence \citep{2013ApJ...775...56K}.  With our analysis showing the orbits of B and C evolve strongly with time, we find repeated passages are far from assured, and thus the mechanism can be effective.  Furthermore, as the time from when a system first looks like Fomalhaut to when it last looks like Fomalhaut is only a few tens of Myrs, the amount of differential procession should be less than would be expected for a system that had been evolving for 440 Myrs.   As seven disks are disrupted, they are incompatible with the observed disk, but represent only a minority of cases.  Thus disruption is a possible but unlikely outcome of the evolutionary model we consider here, making our model compatible with the observed disk around Fomalhaut A.

\begin{figure}
  \centering
  {\includegraphics[width=0.50\textwidth,trim = 120 50 100 50, clip]{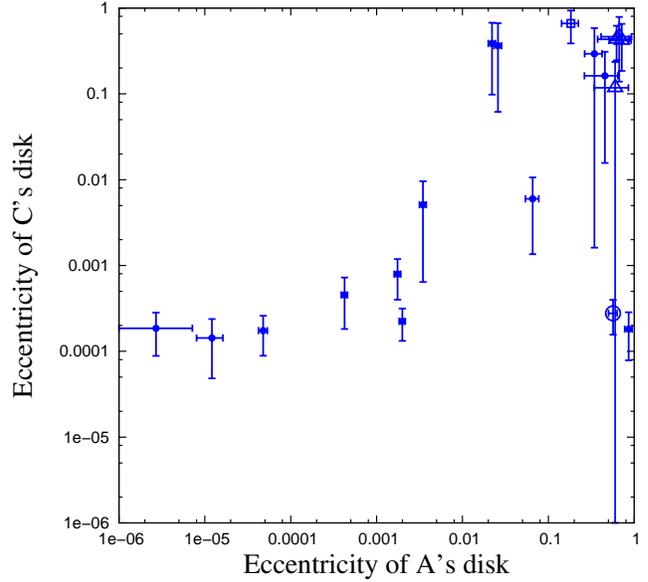}}
  \caption{The mean eccentricty of the disk around A vs. the mean eccentricity of the disk around C, from a subset of 50 simulations performed with debris disks (19 became matches).  The error bars show the standard deviation among the disk particles.  The mean eccentricity of disk particles around A and around C are correlated.  Open symbols denote cases where disk particles were lost (squares where particles were lost from C's disk, circles for A, triangles for both). The correlation between the eccentricity of A's disk and C's disk, along with A's disk's high eccentricity today suggests C's disk may have a high eccentricity, possibly explaining the disk's unusually high brightness.}
\label{fig:diskcomp}
\end{figure}

In figure \ref{fig:diskcomp}, we plot the mean eccentricity of A's disk against the mean eccentricity of C's disk.  We find that the mean eccentricities of the two disks are correlated.  The correlation shows significant scatter, as close encounters with B may be significant source of perturbations to either disk without affecting the other.  Nevertheless, this points to the possibility that the high eccentricity of A's disk may be reflected in a higher disk eccentricity for C than would normally occur for an M star.  Such an externally driven eccentricity might arrest planet formation and initiate a collisional cascade by raising collision velocities \citep[as in][]{2002AJ....123.1757K}, or trigger the instability of a planetary system \citep[as in][]{2004AJ....128..869Z,2011MNRAS.411..859M} which could subsequently stir the disk, leading to the detectability of its debris disk.

\section{Discussion}

Formation of a weakly bound triple system like Fomalhaut is unlikely by capture of two stars during cluster dispersal.  We consider an alternate scenario, where Fomalhaut A and C formed as a tighter binary (with semimajor axis $5000~\rm{au} \lesssim a \lesssim 13600~\rm{au}$)~and Fomalhaut B was captured into a weakly bound orbit during cluster disperal.  We simulate the evolution of such a system, and show that such a system commonly ($55\% - 60\%$) evolves into a Fomalhaut-like system, with both components at large separations from Fomalhaut A, on timescales compatible with the current age of the system.  In this evolution, the present day state where both components are on wide orbits is a temporary one, which typically dissociates in $\mathcal{O}\left(10\right)$~Myrs.  In our simulations, systems that match Fomalhaut may be bound, or else be in the process of ejecting C from the system.  

We perform an additional 50 simulations with disks of test particles around Fomalhaut A and Fomalhaut C, to ascertain whether this formation scenario is compatible with the coherent eccentric disk around Fomalhaut A, and the existence of a detectable disk around Fomalhaut C.  
Nineteen of those simulations became matches for the Fomalhaut system.  In seven cases, the disk around Fomalhaut A is not significantly perturbed, and such cases would be compatible with the dynamics of A's disk being set by an internal source \citep[see e.g.,][]{2006MNRAS.372L..14Q}.  Intriguingly, five systems develop coherent eccentricities in A's disk, owing to secular interactions or close encounters with Fomalhaut C in its much tighter primordial orbit about A, with eccentricities with a factor of 5 of the present day value.  This suggests a possible origin of A's coherently eccentric disk.  We also find that the final excitation of disk eccentricity around A and around C are correlated, suggesting that if A's debris disk was driven to high eccentricity by interactions with C, C's disk may be expected to have been driven to high eccentricity by A.  This may explain why Fomalhaut C has a detectable debris disk, which is rarely found around an M star.

The total likelihood of the scenario presented may seem quite small, when one considers that Fomalhaut is the fourth nearest A star.  However, in matching the details of any system, and focussing on the most interesting aspects, the intrinsic likelihood of that outcome will be small.  The probability that the system we conjectured to be the primordial state of Fomalhaut undergoes an instability to produce a Fomalhaut-like system is $\sim 50\%$.  Although the probability of forming our conjectured initial state is small, $10^5 \rm{au}$~binaries are an order of magnitude less common than $10^4 \rm{au}$ binaries \citep{2010AJ....139.2566D,2014MNRAS.437.1216D}
, and thus given the high chance of creating a Fomalhaut like system from an initially hierarchical system, this should be the preferred model.


There are two observational tests of this scenario. Firstly (as noted above) we find a high likelihood ($> 80\%$) that detailed orbital characterisation of the Fomalhaut system will reveal that C's orbit is such that it interacts strongly with B during pericentre passage. Secondly, further circumstantial evidence in favour
of this scenario would be provided if C were found to possess an eccentric disc. However, this too is not an inevitable consequence of this scenario: Figure \ref{fig:diskcomp} demonstrates that there are situations where A's disc acquires an eccentricity of $\gtrsim 0.1$~while the eccentricity of C's disk remains substantially smaller.

\section{Acknowledgements}
We thank Grant Kennedy and Nick Moeckel for useful discussions, Dimitri Veras for providing us with the code from \citet{2013MNRAS.430..403V}, and the anonymous referee for helpful feedback clarifying this manuscript.  AS and MW are supported by the European Union through ERC grant number 279973.

\bibliography{fomalhaut}

\end{document}